\documentclass[preprintnumbers,eqsecnum,prd,nofootinbib,twocolumn]{revtex4}  

\usepackage{graphicx}
\usepackage{latexsym}
\usepackage{amsmath}
\usepackage{amssymb}
\usepackage{amstext}

\def\beq{\begin{equation}}
\def\eeq{\end{equation}}

\begin{document}

\title{Gravitational self-force corrections to tidal invariants for spinning particles on circular orbits in a Schwarzschild spacetime}

\author{Donato Bini$^1$ and Andrea Geralico$^1$}
  \affiliation{
$^1$Istituto per le Applicazioni del Calcolo ``M. Picone,'' CNR, I-00185 Rome, Italy
}

\date{\today}

\begin{abstract}
We compute gravitational self-force corrections to tidal invariants for spinning particles moving along circular orbits in a Schwarzschild spacetime. In particular, we consider the square and the cube of the gravitoelectric quadrupolar tidal tensor and the square of the gravitomagnetic quadrupolar tidal tensor. 
Our results are accurate to first-order in spin and through the 9.5 post-Newtonian order. We also compute the associated electric-type and magnetic-type eigenvalues.
\end{abstract}


\maketitle

\section{Introduction}

The recent detections of gravitational wave signals by the Ligo and Virgo collaborations \cite{Abbott:2016blz,Abbott:2016nmj,Abbott:2017oio,TheLIGOScientific:2017qsa} have strengthen more and more our motivations and efforts to improve the theoretical understanding of the general relativistic dynamics of binary systems made of compact objects endowed with spin, quadrupole, and even higher multipolar structure. 
In this context tidal interactions are expected to play an important role. For instance, gravitational waves emitted during the coalescence of binary neutron stars are expected to contain an imprint of the tidal interaction between the two bodies \cite{Flanagan:2007ix,Read:2009yp,Baiotti:2010xh,Bernuzzi:2012ci,Damour:2012yf,Bernuzzi:2011aq,Read:2013zra,DelPozzo:2013ala}.

In a zeroth-order approximation level, the mass of one body can be considered much smaller than that of the other, in such a way that backreaction effects can also be neglected and the two-body dynamics reduces to the motion of an extended body in a given gravitational field due to the body of higher mass.
The smaller body thus undergoes tidal deformations which can be studied, e.g., by constructing a body-fixed frame adapted to the timelike geodesic representative of its motion under the action of a tidal potential in terms of Fermi-type coordinates \cite{Ishii:2005xq}. Alternatively, one can use an effective field theory approach to dynamical tides consisting in modifying the point mass action with the addition of certain non-minimal couplings involving integrals of tidal invariant quantities performed along the body's world line \cite{Damour:1995kt,Goldberger:2004jt,Porto:2005ac,Levi:2014gsa,Levi:2015msa,Bini:2015kja}. Such invariants are constructed through the electric-type and magnetic-type tidal tensors given by the tensorial contraction of the spacetime Riemann tensor and its dual, respectively, with the tensor product of the body's four-velocity with itself. 
Actually, such an effective action description of tidal effects is a general framework holding in perturbation theory too, where the backreaction of the body on the background geometry is taken into account, as well as in post-Newtonian theory \cite{Bini:2012gu}.

The mathematical modeling of the two-body tidal problem in complete generality is much complicated due to the impossibility of exploring through analytical methods the strong-field regime (occurring typically at the end of the coalescence process, when the  merging phase starts) and the associated effects, so that one has mainly to rely on numerical analyses. 
Recent years have witnessed a very useful strategy to investigate strong-field effects which combines, in a synergic way, information coming from various  approximation methods, namely: the post-Newtonian (PN) formalism (see, e.g., Ref. \cite{Blanchet:2013haa} and references therein), the post-Minkowskian (PM) one \cite{Blanchet:1984uw,Detweiler:1996mq,Damour:2016gwp}, the gravitational self-force (GSF) formalism (see, e.g., Ref. \cite{Barack:2009ux} and references therein), full numerical relativity simulations \cite{Hotokezaka:2013mm,Radice:2013hxh,Bernuzzi:2014kca}, and the effective one-body (EOB) formalism \cite{Buonanno:1998gg,Buonanno:2000ef,Damour:2000we,Damour:2001tu}. 	
In particular, the EOB approach has proven to be a key framework where one can combine efficiently information coming from all the other approximation schemes, with the advantage that EOB-based simulations are very fast (and accurate), and thousands of EOB templates have been built over the last years \cite{Taracchini:2013rva,Purrer:2015tud}.
At the moment, the inclusion in the EOB Hamiltonian of tidal interaction terms is still under study.
For instance, Ref. \cite{Steinhoff:2016rfi} considers finite size effects on the orbital dynamics of a compact body like a neutron star by modifying the (effective) point-particle action adding quadrupolar degrees of freedom, which in turn allow to define \lq\lq dynamical tides'' all along the inspiral process. 

In the present paper, we extend previous works on tidal invariants along circular orbits in a Schwarzschild spacetime obtained for spinless particles \cite{Bini:2012gu,Bini:2014zxa,Dolan:2014pja,Kavanagh:2015lva,Nolan:2015vpa,Bini:2015kja,Shah:2015nva} to the case of spinning bodies to linear order in spin in the extreme-mass-ratio limit. 
This work belongs to a project studying gravitational self-force corrections due to the multipolar structure of the perturbing body, already started with the computation of the Detweiler's redshift invariant \cite{Bini:2018zde}.
As standard, the dynamics of the small spinning body is described according to the Mathisson-Papapetrou-Dixon (MPD) model \cite{Mathisson:1937zz,Papapetrou:1951pa,Dixon:1970zza}. The orbit is assumed to be circular and equatorial, the spin vector being orthogonal to the plane of motion. 

We expect that the present work will give additional (high-PN order, first-GSF order) information to be used to improve both modeling and simulations of the two-body dynamics in the extreme mass ratio and small spin limits.
 
The masses of the gravitationally interacting two bodies are denoted by $m_1$ and $m_2$, with the convention that $m_1\le m_2$.
We then define
\begin{eqnarray}
\label{eq:1.1}
M\equiv m_1+m_2\,,\quad \mu\equiv\frac{m_1 m_2}{M}\,,\quad
\nu \equiv \frac{\mu}{M}\,,
\end{eqnarray}
as the total mass, reduced mass and {\it symmetric} mass ratio, respectively.
We shall also use the other dimensionless mass ratios
\beq
\label{eq:1.2}
q\equiv \frac{m_1}{m_2}\,,\quad X_1\equiv \frac{m_1}{M}\,, \quad X_2\equiv \frac{m_2}{M}= 1-X_1 
\,,
\eeq
such that
\beq
\label{eq:1.3}
\nu=\frac{q}{(1+q)^2}\,,\quad X_1=\frac{q}{1+q}\,, \quad
X_2=\frac{1}{1+q}\,.
\eeq
In the small mass-ratio limit $q\ll 1$ we have $\nu \simeq X_1\simeq q\ll 1$ and $X_2=1-q+O(q^2)$.

In the following we will set $G=c=1$ and use the signature $+2$.

\section{Spinning particle motion in a perturbed Schwarzschild spacetime}

Let us consider a spinning particle with mass $m_1$ moving in a perturbed Schwarzschild spacetime with line element 
\beq
ds^2=(\bar g_{\alpha\beta} +h_{\alpha\beta})dx^\alpha dx^\beta \,,
\eeq
where 
\begin{eqnarray}
\bar g_{\alpha\beta}dx^\alpha dx^\beta
&=&-fdt^2+f^{-1}dr^2+r^2(d\theta^2+\sin^2\theta d\phi^2)\,,\nonumber\\
\end{eqnarray}
with $f=1-{2m_2}/{r}$, and $h_{\alpha\beta}=O(q)$.
The motion is governed by the MPD equations \cite{Mathisson:1937zz,Papapetrou:1951pa,Dixon:1970zza}, which read 
\begin{eqnarray}
\label{papcoreqs1}
m_1 \frac{{\rm D}U^{\mu}}{d \tau} & = &
- \frac12 \, R^\mu{}_{\nu \alpha \beta} U^\nu S^{\alpha \beta} \,,
\\
\label{papcoreqs2}
\frac{{\rm D}S^{\mu\nu}}{d \tau} & = & O(2)\,,
\end{eqnarray}
to first-order in spin.
Here we recall that $U^{\mu} \equiv d z^\mu/d \tau$ is the timelike unit tangent vector to the world line representative of the body's motion (with parametric equations $x^\mu=z^\mu(\tau)$) used to make the multipole reduction, parametrized by the proper time $\tau$, and the total 4-momentum of the particle is aligned with $U$, i.e., $P=m_1 U+O(2)$.
The (antisymmetric) spin tensor $S^{\mu \nu}$ is taken to satisfy the Tulczyjew-Dixon conditions~\cite{Dixon:1970zza,tulc59} 
\beq
\label{tulczconds}
S^{\mu\nu} P_{\nu} = m_1 \, S^{\mu\nu}U{}_\nu=0\,.
\eeq
The associated spin vector obtained by spatial duality
\beq
S(U)^\gamma 
=\frac12 U_\sigma \eta^{\sigma\gamma\alpha\beta}S_{\alpha\beta}\,,\qquad S^\gamma U_\gamma=0\,,
\eeq
is parallel-propagated along $U$, $\nabla_U S^\gamma=0$.
Both the particle's mass $m_1$ and the signed magnitude $s$ of the spin vector
\beq
\label{sinv}
s^2=S(U)^\beta S(U)_\beta = \frac12 S_{\mu\nu}S^{\mu\nu}\,, 
\eeq
are constant along the path.

Let us assume that the perturbed metric admits the Killing vector $k=\partial_t+\Omega\partial_\phi$ and that the body's orbit is aligned with $k$ with spin vector directed orthogonal to the equatorial plane, i.e., it moves along a circular orbit $U=\Gamma k$ with
\beq
\label{Gamma_def}
-\Gamma^{-2}=k\cdot k=-f +\Omega^2 r^2 +h_{kk}\,,
\eeq
where $h_{kk}=h_{\alpha\beta}k^\mu k^\nu$, and
\beq
S=-s e_{\hat \theta}\,,\qquad
e_{\hat \theta}=\frac1r\left(1-\frac1{2r^2}h_{\theta\theta}\right)\partial_\theta\,.
\eeq
The spin magnitude $s$ has a positive (negative) sign
if the associated orbital angular momentum is parallel (respectively, antiparallel) to $e_{\hat z}=-e_{\hat \theta}$.
It is also useful to introduce the associated dimensionless spin parameter $\hat s\equiv s/(m_1m_2)$. 

The equations of motion \eqref{papcoreqs1} can then be cast in the form
\begin{eqnarray}
\label{eq_moto_k}
m_1 \Gamma  \nabla_k k^{\mu}&=&
- \frac12 ( \nabla_{\mu\beta}k_\alpha) \, S^{\alpha \beta}\nonumber\\
&=&- \frac12  (\nabla_{\mu}K_{\beta\alpha} )\, S^{\alpha \beta}\,, 
\end{eqnarray}
where  the (antisymmetric) tensor $K_{\alpha\beta}$ is given by
\beq
K_{\alpha\beta}=\nabla_\alpha k_\beta=\partial_{[\alpha}k_{\beta]}\,,
\eeq
and imply the following solution for $\Omega$
\beq
m_2\Omega=u^{3/2}\left[
1-\frac32{\hat s}u^{3/2}+q(\tilde\Omega_1+{\hat s}\tilde\Omega_{1{\hat s}})
\right]\,.
\eeq
Here 
\beq
\label{tildeOm1def}
\tilde\Omega_1=
-\frac{m_2}{4u^2}\partial_r h_{kk}^{(0)}\,,
\eeq
and the spin correction
\beq
\label{tildeOm1sdef}
\tilde\Omega_{1{\hat s}}=\tilde\Omega_{1{\hat s}}^{(h)}+\tilde\Omega_{1{\hat s}}^{(\partial h)}+\tilde\Omega_{1{\hat s}}^{(\partial^2 h)}\,,
\eeq
with
\begin{widetext}
\begin{eqnarray}
\tilde\Omega_{1{\hat s}}^{(h)}&=&
-\frac{u^{3/2}}{4(1-2u)^2}h_{kk}^{(0)}
+\frac{(5-12u)u^{3/2}}{4}h_{rr}^{(0)}
-\frac{u^2(3-4u)(1-3u)}{2m_2(1-2u)^2}h_{t\phi}^{(0)}
-\frac{(1-3u)(2-5u+4u^2)u^{5/2}}{4m_2^2(1-2u)^2}h_{\phi\phi}^{(0)}
\,,\nonumber\\
\tilde\Omega_{1{\hat s}}^{(\partial h)}&=&
\frac{m_2u^{1/2}}{4(1-2u)}\partial_r h_{kk}^{(0)}
-\frac{m_2}{4u^2}\partial_r h_{kk}^{(1)}\nonumber\\
&&
+\frac14(1-3u)\left[
-\partial_{\phi} h_{rk}^{(0)}
+m_2(1-2u)u^{-1/2}\partial_r h_{rr}^{(0)}
+\frac{1}{(1-2u)}\partial_r h_{t\phi}^{(0)}
+\frac{(2-3u)u^{3/2}}{m_2(1-2u)}\partial_r h_{\phi\phi}^{(0)}
\right]
\,,\nonumber\\
\tilde\Omega_{1{\hat s}}^{(\partial^2 h)}&=&
-\frac{m_2^2u^{-3/2}}{4}\partial_{rr} h_{kk}^{(0)}
-\frac{m_2}{4u}(1-3u)[\partial_{rr} h_{\phi k}^{(0)}-\partial_{r\phi} h_{rk}^{(0)}]
\,,
\end{eqnarray}
\end{widetext}
where the metric components $h_{\alpha\beta}=h_{\alpha\beta}^{(0)}+\hat s h_{\alpha\beta}^{(1)}$ and their derivatives are evaluated at $r=m_2/u$.

It is useful to introduce the dimensionless frequency parameter $y=(m_2\Omega)^{2/3}$,
\beq
y=u-{\hat s}u^{5/2}+q {\mathcal F}(u)\,, 
\eeq
to first order in $\hat s$, where ${\mathcal F}(u)={\mathcal F}_0(u)+\hat s {\mathcal F}_{\hat s}(u)$.
A direct calculation shows
\begin{eqnarray}
{\mathcal F}_0(u)&=&\frac23u\tilde\Omega_1(u)\,,\nonumber\\
{\mathcal F}_{\hat s}(u)&=&\frac13u^{5/2}\tilde\Omega_1(u)+\frac23u\tilde\Omega_{1{\hat s}}(u)\,.
\end{eqnarray}
This relation can be inverted to give
\beq
u=y\left(1+{\hat s}y^{3/2}\right)+q [f_0(y)+\hat s f_{\hat s}(y)]\,, 
\eeq
to first order in $\hat s$, where 
\begin{eqnarray}
f_0(y)&=&\frac{m_2}{6y}\partial_r h_{kk}^{(0)}\,,\nonumber\\
f_{\hat s}(y)&=&\frac{y}{3}\left[
\frac{m_2}{\sqrt{y}}\partial_r h_{kk}^{(0)}-2\tilde\Omega_{1{\hat s}}
\right]\,.
\end{eqnarray}

Finally, the following quantity is a constant of motion associated with the Killing vector $k$
\begin{eqnarray}
E-\Omega J&=&-m_1 k_\alpha U^\alpha+\frac12S^{\alpha\beta}\nabla_\beta k_\alpha\nonumber\\
&=&\frac{m_1}{\Gamma}+\Gamma K^*{}^{\sigma\gamma}\, k_\sigma S_\gamma\nonumber\\
&=& m_1 \Gamma^{-1}-m_1\hat sm_2b
\,,
\end{eqnarray}
so that 
\beq
\hat E-m_2\Omega \hat J= z_1-\hat sm_2b\,,
\eeq
with $\hat E=E/m_1$, $\hat J=J/(m_1 m_2)$ and $z_1=\Gamma^{-1}$.
The quantity
\beq
m_2b=u^{3/2}(1-3u)^{1/2}(1+q\delta_b(u))\,,
\eeq
with
\begin{eqnarray}
\delta_b(u)&=&
-\frac{m_2(1-4u)}{4u^2(1-3u)}\partial_r h_{kk}^{(0)}
+\frac{1}{u^{1/2}}\,\partial_{[r} h_{\phi]k}^{(0)}\nonumber\\
&-&
\frac12(1-2u) h_{rr}^{(0)}
-\frac{u}{2(1-3u)(1-2u)}h_{kk}^{(0)}\nonumber\\
&-&
\frac{u^{3/2}}{m_2(1-2u)} \left(h_{t\phi}^{(0)}
+\frac{u^{1/2}(1-u)}{2m_2} h_{\phi\phi}^{(0)}\right)
\,,\nonumber\\
\end{eqnarray}
is related to the spin precession frequency.
In fact, the previous equation can be rewritten as
\beq
\delta_b(u)=-\frac{m_2(1-4u)}{4u^2(1-3u)}\partial_r h_{kk}^{(0)}-\delta(u)
\,,
\eeq
where $\delta(u)$ was defined in Eq. (3.9) of Ref. \cite{Bini:2014ica}.
Passing to the variable $y$, the quantity $\delta_b(y)$ simply becomes $\delta_b(y)=-\delta(y)$.
A direct comparison with Eq. (4.8) of Ref. \cite{Blanchet:2012at} then leads to the identification $b=\Omega-\Omega_1$, where $\Omega$ is the orbital frequency and $\Omega_1$ the precession frequency.
It is also related to the spin precession angle $\psi$ by $b=\Omega\psi$ (see Ref. \cite{Dolan:2013roa}).

\section{GSF corrections to tidal invariants in the presence of spin}

In any given spacetime the electric-tidal forces (relative to $U$) are governed by the \lq\lq potentials"
\begin{eqnarray}
{\rm Tr}[{\mathcal E}(U)^2]&=&{\mathcal E}(U)_{\alpha\beta}{\mathcal E}(U)^{\alpha\beta}\,,\nonumber\\
{\rm Tr}[{\mathcal E}(U)^3]&=&{\mathcal E}(U)_{\alpha\beta}{\mathcal E}(U)^{\beta\mu}{\mathcal E}(U)_{\mu}{}^{\alpha}\,,
\end{eqnarray}
where
\beq
{\mathcal E}(U)_{\alpha\beta}=R_{\alpha\mu\beta\nu}U^\mu U^\nu\,.
\eeq
Similarly, the magnetic-tidal forces are governed by the potential
\beq
{\rm Tr}[{\mathcal B}(U)^2]={\mathcal B}(U)_{\alpha\beta}{\mathcal B}(U)^{\alpha\beta}\,,
\eeq 
where
\beq
{\mathcal B}(U)_{\alpha\beta}=R^*{}_{\alpha\mu\beta\nu}U^\mu U^\nu\,.
\eeq
Here $U=\Gamma k$ (with $\Gamma=U^t$) is the four velocity of the spinning particle and is aligned with a  nongeodesic orbit (differently from previous works) because of the nonvanishing spin of the particle which implies the existence of spin-curvature coupling forces. Expanding $\Gamma$ and $\Omega$ in powers of $\hat s$ we find (at the first order in $\hat s$)
\begin{eqnarray}
\Gamma &=&\bar\Gamma +q (\Gamma_{1} +\hat s \Gamma_{1\hat s})\,,\nonumber\\
\Omega &=&\bar\Omega +q (\Omega_{1} +\hat s \Omega_{1\hat s})\,,
\end{eqnarray}
where
\begin{eqnarray}
\bar\Gamma&=&\frac{1}{\sqrt{1-3u}} \left(1- \frac32{\hat s}  \frac{u^{5/2}}{1-3u}\right)\,,\nonumber\\
m_2\bar\Omega&=&u^{3/2}\left(1-\frac32{\hat s}u^{3/2}\right)\,.
\end{eqnarray}
The first-order self-force (1SF) corrections to the frequency $\Omega_{1}=u^{3/2}\tilde\Omega_{1}$ and $\Omega_{1\hat s}=u^{3/2}\tilde\Omega_{1\hat s}$ are given in Eqs. \eqref{tildeOm1def}--\eqref{tildeOm1sdef} above, whereas the correction to the redshift factor $\Gamma=U^t=1/z_1$ can be found in Ref. \cite{Bini:2018zde}.

Let us consider the electric and magnetic part of the Riemann tensor with respect to $k$ instead of $U$ \cite{Bini:2014zxa}.
We compute both the quadratic tidal-electric and tidal-magnetic invariants as well as the cubic tidal-electric invariant defined by 
\begin{eqnarray}
{\mathcal J}_{e^2}&\equiv&{\rm Tr}[{\mathcal E}^2(k)]={\mathcal E}^\mu{}_\nu(k){\mathcal E}^\nu{}_\mu(k)\,,\nonumber\\
{\mathcal J}_{b^2}&\equiv&{\rm Tr}[{\mathcal B}^2(k)]={\mathcal B}^\mu{}_\nu(k){\mathcal B}^\nu{}_\mu(k)\,,\nonumber\\
{\mathcal J}_{e^3}&\equiv&{\rm Tr}[{\mathcal E}^3(k)]={\mathcal E}^\mu{}_\nu(k){\mathcal E}^\nu{}_\rho(k){\mathcal E}^\rho{}_\mu(k)\,,
\end{eqnarray}
respectively, where 
\begin{eqnarray}
{\mathcal E}_{\alpha\beta}(k)&\equiv& R_{\alpha\mu \beta \nu} k^\mu k^\nu
= \Gamma^{-2}{\mathcal E}_{\alpha\beta}(U)\,,\nonumber\\
{\mathcal B}_{\alpha\beta}(k)&\equiv& R^*_{\alpha\mu \beta \nu}k^\mu k^\nu
= \Gamma^{-2}{\mathcal B}_{\alpha\beta}(U)\,.
\end{eqnarray}

The 1SF-accurate expansion of the $m_2$-adimensionalized version of both electric-like and magnetic-like tidal invariants, expressed in terms of $y$, reads
\begin{eqnarray}
\tilde {\mathcal J}_{e^2}(y)&\equiv&m_2^4 {\mathcal J}_{e^2}
=\tilde {\mathcal J}_{e^2}^{(0)}[1+q\, \delta_{e^2}(y)+O(q^2)]\,,\nonumber\\
\tilde {\mathcal J}_{b^2}(y)&\equiv&m_2^4{\mathcal J}_{b^2}
=\tilde {\mathcal J}_{b^2}^{(0)}[1+q \delta_{b^2}(y)+O(q^2)]\,, \nonumber\\
\tilde {\mathcal J}_{e^3}(y)&\equiv&m_2^6{\mathcal J}_{e^3}
=\tilde {\mathcal J}_{e^3}^{(0)}[1+q \delta_{e^3}(y) +O(q^2)]\,,
\end{eqnarray}
where 
\begin{eqnarray}
\tilde {\mathcal J}_{e^2}^{(0)}&=&
6y^6(1-3y+3y^2)+36y^{15/2}(1-2y)^2\hat s\,,\nonumber\\
\tilde {\mathcal J}_{b^2}^{(0)}&=&
18y^7(1-2y)+36y^{17/2}(2-5y)\hat s\,,\nonumber\\
\tilde {\mathcal J}_{e^3}^{(0)}&=&
-3y^9(1-3y)(2-3y)\nonumber\\
&&
-27y^{21/2}(1-2y)(1-3y)(2-y)\hat s\,.
\end{eqnarray}

The GSF corrections $\delta_{e^2}$, $\delta_{b^2}$ and $\delta_{e^3}$ are understood to be regularized following the standard GSF procedure (which we do not need to repeat here) by subtracting their singular parts (here the \lq\lq $B$-term").\footnote{
Actually the computation of the Detweiler-Whiting full singular field would require additional regularization parameters (see, e.g., Ref. \cite{Heffernan:2012su} and references therein). We only subtract the $B$-term which is necessary to have a convergent series. 
}
In this way one is left with convergent series, e.g.,  
\beq
\delta^{\rm reg}=\sum_{l=0}^\infty \left( \delta_l^0-B(y;l)\right)\,,
\eeq
where 
\beq
\delta_l^0\equiv \frac12 (\delta_l^+ + \delta_l^-)\,;
\eeq
the \lq\lq subtraction term'' or \lq\lq B-term" is of the form
\beq
B(y;l)=l(l+1)b_0(y)+b_1(y)\,,
\eeq
with $b_0(y)=b_0^{\hat s^0}(y)+\hat s b_0^{\hat s^1}(y)$ and similarly for $b_1(y)$.

To make a long story short, we expand the metric perturbations in tensorial spherical harmonics in the Regge-Wheeler gauge \cite{Regge:1957td} following the original approach of Zerilli \cite{Zerilli:1971wd}. We solve then the resulting radial equations in PN-sense and use up to a fixed PN-order. To reach the latter order is necessary to combine  PN type solutions with Mano-Suzuki-Takasugi (MST) type  solutions \cite{Mano:1996mf,Mano:1996vt}, following a standard approach first introduced in Ref. \cite{Bini:2013zaa}.

We will use here the MST solutions for the multipoles $l=2,\ldots,7$, so that our result will be accurate up to the order $O(y^{19/2})$ included, corresponding to the 9.5PN order.

\subsection{Tidal-electric: ${\rm Tr}[{\mathcal E}^2(k)]$}

The $O(q)$ perturbation to ${\rm Tr}[{\mathcal E}^2(k)]$ is given by the following combination of perturbed metric components $h_{\mu\nu}$ and their derivatives
\begin{eqnarray}
\delta_{e^2}(y)&=&\delta_{e^2, h}(y)+\delta_{e^2, \partial h}(y)+\delta_{e^2, \partial^2 h}(y)\,, 
\end{eqnarray}
where

\begin{widetext}

\begin{eqnarray}
\delta_{e^2, h}(y)&=&  
\left[-\frac{y}{3 (-1+2 y) (1-3 y+3 y^2)} +\hat s\frac{ y^{3/2} (15y^3-105y^4+90y^5+30y^2-17y+3)}{3(-1+2 y)^2 (1-3 y+3 y^2)^2} \right] h_{kk}\nonumber\\
&+&\left[\frac{ (-1+2 y) (18 y^2-18 y+5)}{3(1-3 y+3 y^2)}+\hat s\frac{ (594y^5-1341y^4+1293y^3-657y^2+172y-18) y^{3/2}}{3(1-3 y+3 y^2)^2} \right] h_{rr}\nonumber\\
&+& \left[ -\frac{y^2(3y-1)(y-1)}{3(-1+2 y)(1-3 y+3 y^2) M^2}\right.\nonumber\\
&&\left.
-\hat s\frac{ (3033y^5-1989y^6+594y^7-2643y^4+1397y^3-445y^2+79y-6) y^{5/2}}{3 M^2 (-1+2y)^2(1-3 y+3 y^2)^2} \right] h_{\phi\phi}\nonumber\\
&+&\left[-\frac{ y^2}{3(1-3 y+3 y^2) M^2} +\hat s\frac{2 (18 y^3-18 y^2+6 y-1) y^{7/2}}{3 M^2 (1-3 y+3 y^2)^2}\right] h_{\theta\theta}\nonumber\\
&+&\left[\frac{2 y^{3/2}(3y-1)}{3 M (-1+2y)(1-3 y+3 y^2)} -\hat s\frac{4 (3 y-2) y^5}{(1-3 y+3 y^2)^2 M} \right] h_{t\phi}\nonumber\\
\delta_{e^2, \partial h}(y)&=&   
\left[\frac{M(3y-1)(3y-2)}{3y^2(1-3 y+3 y^2)}+\hat s\frac{M(225y^2-333y^3+225y^4+10-78y)}{6y^{1/2}(1-3y+3y^2)^2}\right]\partial_r h_{kk} \nonumber\\
&+&\left[\frac{(3y-1)y}{3 M (1-3 y+3 y^2)} +\hat s \frac{y^{3/2}(243y^5-783y^4+888y^3-471y^2+121y-12)}{6(1-3 y+3 y^2)^2 M}  \right] \partial_r h_{\phi\phi} \nonumber\\
&+&\left[\frac{(3y-1)}{ 3y^{1/2}(1-3 y+3 y^2)}-\hat s\frac{(3y^2+3y-1)(27y^3-36y^2+17y-3)}{3(1-3 y+3 y^2)^2}\right]\partial_r h_{t\phi}  \nonumber\\
&+&\left[-\frac{(3y-1)y}{3M(1-3y+3y^2)} -\hat s\frac{y^{3/2}(6y^2-6y+1)(18y^3-33y^2+17y-3)}{3M(1-3y+3y^2)^2} \right]\partial_\phi h_{r\phi}  \nonumber\\
&+&\left[-\frac{(3y-1)}{3y^{1/2}(1-3y+3y^2)}-\hat s\frac{(27y^5-171y^4+228y^3-126y^2+32y-3)}{3(1-3y+3y^2)^2}\right]\partial_\phi h_{tr} \nonumber\\
&+& \hat s\frac{(-1+2y)y^{5/2}}{M(1-3y+3y^2)}(\partial_\theta h_{r\theta }-\partial_r h_{\theta\theta}) \nonumber\\
&-&\hat s\frac{(-1+2y)^2(15y^2-12y+2)}{2y^{1/2}(1-3y+3y^2)}\partial_r h_{rr}\nonumber\\
\delta_{e^2, \partial^2 h}(y)&=&  
\left[\frac{(-1+2y)(3y-2)M^2}{6 y^3 (1-3 y+3 y^2)} +\hat  s\frac{M^2(-1+9y-33y^2+36y^3)}{6y^{1/2} (1-3 y+3 y^2)^2}\right]\partial_{rr} h_{kk}  \nonumber\\
&+&\left[-\frac{1}{6(-3y+1+3y^2)y}+\hat s\frac{(18y^3-9y^2-3y+1)y^{1/2}}{6(-3y+1+3y^2)^2}\right]\partial_{\theta\theta} h_{kk} \nonumber\\
&+&\left[\frac{(3y-1)^2}{6(-3y+1+3y^2)(-1+2y)y}+\hat s\frac{y^{1/2}(24y^3-33y^2+13y-1)(3y-1)^2}{6(-1+2y)^2(-3y+1+3y^2)^2}\right]\partial_{\phi\phi} h_{kk}  \nonumber\\
&-&\hat s\left[\frac{(3y-1)(-1+2y)^2M}{(-3y+1+3y^2)y}\right](\partial_{rr} h_{\phi k}-\partial_{r\phi} h_{rk} )\,.
\end{eqnarray}

The regularized value of the zeroth order in spin correction $\delta_{e^2}^{\hat s^0}(y)$ is known \cite{Bini:2014zxa,Dolan:2014pja,Kavanagh:2015lva,Nolan:2015vpa}.
We recall below for completeness only the first few PN terms
\begin{eqnarray}
\delta_{e^2}^{\hat s^0}(y) &=& -2+5 y+\frac{61}{4} y^2+\left(\frac{593}{256}\pi^2-\frac{1669}{24}\right) y^3\nonumber\\
&+& \left(\frac{1118879}{4800}-\frac{5867}{1024}\pi^2-\frac{768}{5}\gamma-\frac{1536}{5} \ln(2)-\frac{384}{5} \ln(y)\right) y^4
+O_{{\rm ln}}(y^5)\,.
\end{eqnarray}
The correction linear in spin is instead new and given by 

\begin{eqnarray}
\delta_{e^2}^{\hat s^1}(y) &=&
-2 y^{3/2}+15 y^{5/2}-\frac{107}{4}y^{7/2}+\left(-\frac{103}{24}+\frac{41}{16}\pi^2\right) y^{9/2}\nonumber\\
&+&\left(-\frac{292027}{576}+\frac{68867}{1536}\pi^2+\frac{1576}{15}\ln(y)+432\ln(2)+\frac{3152}{15}\gamma\right) y^{11/2}\nonumber\\
&+&\left(-\frac{6751671}{3200}-\frac{42008}{35}\ln(2)+\frac{3764}{35}\ln(y)+\frac{7528}{35}\gamma+\frac{10935}{7}\ln(3)+\frac{172241}{8192}\pi^2\right) y^{13/2}\nonumber\\
&+&\frac{7704}{35}\pi y^7\nonumber\\
&+& \left(\frac{771338065013}{10160640}-\frac{6674991929}{884736}\pi^2-\frac{1037367}{70}\ln(3)+\frac{4392574}{567}\ln(2)-\frac{19876714}{2835}\gamma-\frac{9938357}{2835}\ln(y)\right.\nonumber\\
&+& \left.\frac{8432821}{131072}\pi^4\right) y^{15/2}\nonumber\\
&+& \frac{166681}{2205}\pi y^8\nonumber\\
&+&\left(\frac{703288938133497911}{1760330880000}+\frac{2606489917}{1212750}\ln(y)+\frac{2606489917}{606375}\gamma+\frac{491929443}{12320}\ln(3)+\frac{712890625}{28512}\ln(5)\right.\nonumber\\
&&-\frac{766912112971}{5457375}\ln(2)-\frac{11357408}{1575}\ln(2)^2-\frac{2811104}{1575}\gamma^2-\frac{702776}{157}5\ln(y)^2+\frac{52544}{15}\zeta(3)\nonumber\\
&&-\frac{844288199909}{402653184}\pi^4-\frac {74442851762681}{4954521600}\pi^2-\frac{3773248}{525}\gamma\ln(2)\nonumber\\
&&\left. -\frac{1886624}{525}\ln(2)\ln(y)-\frac{2811104}{1575}\gamma\ln(y)\right) y^{17/2}\nonumber\\
&-&\frac{4923636673}{727650}\pi y^9\nonumber\\
&+&\left(\frac{22417444716344159237593}{28193459374080000}+\frac{26271208759907}{29429400}\ln(2)+\frac{11228736392577}{35672000}\ln(3)\right.\nonumber\\
&&+\frac{159032483612257}{1589187600}\ln(y)+\frac{159479998840417}{794593800}\gamma-\frac{852930}{49}\ln(3)^2-\frac{646076171875}{1729728}\ln(5)\nonumber\\
&& +\frac{303921248}{11025}\ln(2)^2-\frac{69102496}{11025}\gamma^2-\frac{17275624}{11025}\ln(y)^2\nonumber\\
&&+\frac{237056}{7}\zeta(3)-\frac{5973531733489}{536870912}\pi^4-\frac{948482819940703}{277453209600}\pi^2+\frac{110477504}{11025}\gamma\ln(2)\nonumber\\
&& +\frac{55238752}{11025}\ln(2)\ln(y)-\frac{69102496}{11025}\gamma\ln(y)-\frac{1705860}{49}\gamma\ln(3)-\frac{1705860}{49}\ln(2)\ln(3)\nonumber\\
&&\left. -\frac{852930}{49}\ln(3)\ln(y)\right) y^{19/2}
+O_{\ln{}}(y^{10})\,.
\end{eqnarray}

\subsection{Tidal-electric-cube: ${\rm Tr}[{\mathcal E}^3(k)]$}

The $O(q)$ perturbation to ${\rm Tr}[{\mathcal E}^3(k)]$ is given by $\delta_{e^3}(y)$ which has a formal expression in terms of the metric components similar to $\delta_{e^2}(y)$.
The regularized value of the zeroth order in spin correction $\delta_{e^3}^{\hat s^0}(y)$ is known \cite{Bini:2014zxa,Dolan:2014pja,Kavanagh:2015lva,Nolan:2015vpa}.
We recall below for completeness only the first few PN terms
\begin{eqnarray}
\delta_{e^3}^{\hat s^0}(y) &=& -3+\frac{15}{2}y+\frac{147}{8}y^2+\left(-\frac{1561}{16}+\frac{1779}{512}\pi^2\right)y^3\nonumber\\
&&
+\left(\frac{1336679}{3200}-\frac{2271}{256}\pi^2-\frac{1152}{5}\gamma-\frac{2304}{5}\ln(2)-\frac{576}{5}\ln(y)\right)y^4
+O_{{\rm ln}}(y^5)\,.
\end{eqnarray}
The correction linear in spin is instead new and given by 

\begin{eqnarray}
\delta_{e^3}^{\hat s^1}(y) &=& 
-3y^{3/2}+\frac{45}{2}y^{5/2}-\frac{105}{8}y^{7/2}+\left(-\frac{535}{16}+\frac{123}{32}\pi^2\right)y^{9/2}\nonumber\\
&&
+\left(-\frac{371947}{384}+\frac{138301}{2048}\pi^2+\frac{1576}{5}\gamma+648\ln(2)+\frac{788}{5}\ln(y)\right)y^{11/2}\nonumber\\
&&
+\left(-\frac{34633509}{6400}+\frac{702285}{4096}\pi^2+\frac{34224}{35}\gamma-\frac{17232}{35}\ln(2)+\frac{17112}{35}\ln(y)+\frac{32805}{14}\ln(3)\right)y^{13/2}\nonumber\\
&&
+\frac{11556}{35}\pi y^7\nonumber\\
&&
+\left(\frac{2406372381697}{33868800}+\frac{2274533}{189}\ln(2)-\frac{1310013}{70}\ln(3)-\frac{8095607}{1890}\ln(y)-\frac{8095607}{945}\gamma-\frac{9113917597}{1179648}\pi^2\right.\nonumber\\
&&\left.
+\frac{25298463}{262144}\pi^4\right)y^{15/2}\nonumber\\
&&+\frac{1146373}{1470}\pi y^8\nonumber\\
&&+\left(\frac{594053200308222311}{1173553920000}-\frac{1886624}{175}\gamma\ln(2)-\frac{943312}{175}\ln(2)\ln(y)-\frac{1405552}{525}\gamma\ln(y)-\frac{824612241077}{268435456}\pi^4\right.\nonumber\\
&&-\frac{22341281351653}{1651507200}\pi^2-\frac{1405552}{525}\gamma^2-\frac{5678704}{525}\ln(2)^2-\frac{351388}{525}\ln(y)^2+\frac{26272}{5}\zeta(3)-\frac{369489776273}{1819125}\ln(2)\nonumber\\
&& \left.-\frac{1443347929}{404250}\ln(y)-\frac{1443347929}{202125}\gamma+\frac{712890625}{19008}\ln(5)+\frac{976942377}{24640}\ln(3)\right) y^{17/2}\nonumber\\
&& -\frac{2053395569}{242550}\pi y^9\nonumber\\
&&+\left(\frac{49851744177689427024889}{18795639582720000}-\frac{23028752}{3675}\gamma\ln(2)-\frac{11514376}{3675}\ln(2)\ln(y)-\frac{54127112}{3675}\gamma\ln(y)\right.\nonumber\\
&&-\frac{2558790}{49}\gamma\ln(3)-\frac{2558790}{49}\ln(2)\ln(3)-\frac{1279395}{49}\ln(3)\ln(y)-\frac{9399098116257}{335544320}\pi^4-\frac{11234318687460683}{369937612800}\pi^2\nonumber\\
&&-\frac{54127112}{3675}\gamma^2+\frac{73711096}{3675}\ln(2)^2-\frac{13531778}{3675}\ln(y)^2+\frac{2143824}{35}\zeta(3)+\frac{1000473336848339}{882882000}\ln(2)\nonumber\\
&&  +\frac{759486010098089}{5297292000}\ln(y)+\frac{761723586238889}{2648646000}\gamma-\frac{585646484375}{1153152}\ln(5)-\frac{1279395}{49}\ln(3)^2\nonumber\\
&&\left. +\frac{370377504565641}{784784000}\ln(3)\right) y^{19/2}
+O_{\ln{}}(y^{10})\,.
\end{eqnarray}

\subsection{Tidal-magnetic: ${\rm Tr}[{\mathcal B}^2(k)]$}

The $O(q)$ perturbation to ${\rm Tr}[{\mathcal B}^2(k)]$ is given by $\delta_{b^2}(y)$ which has a formal expression in terms of the metric components similar to $\delta_{e^2}(y)$.
The regularized value of the zeroth order in spin correction $\delta_{b^2}^{\hat s^0}(y)$ is known \cite{Bini:2014zxa,Dolan:2014pja,Kavanagh:2015lva,Nolan:2015vpa}.
We recall below for completeness only the first few PN terms
\begin{eqnarray}
\delta_{b^2}^{\hat s^0}(y) &=& -\frac43+\frac{14}{3}y-\frac{11}{6}y^2+\left(-\frac{1723}{36}+\frac{41}{24}\pi^2\right) y^3\nonumber\\
&&+\left(-\frac{357079}{4320}+\frac{73559}{4608}\pi^2-\frac{3616}{45}\gamma-160\ln(2)-\frac{1808}{45}\ln(y)\right) y^4
+O_{{\rm ln}}(y^5)\,.
\end{eqnarray}
The correction linear in spin is instead new and given by

\begin{eqnarray}
\delta_{b^2}^{\hat s^1}(y) &=& 
-\frac{4}{3}y^{3/2}+\frac{37}{3}y^{5/2}+\left(-19+\frac{63}{1024}\pi^2\right) y^{7/2}\nonumber\\
&&+\left(-\frac{62711}{1800}+\frac{1081}{768}\pi^2-\frac{8}{15}\ln(y)-\frac{16}{15}\ln(2)-\frac{16}{15}\gamma\right) y^{9/2}\nonumber\\
&&+\left(-\frac{222888751}{151200}+\frac{87754979}{589824}\pi^2+\frac{2392}{45}\gamma+\frac{6728}{63}\ln(2)+\frac{1196}{45}\ln(y)\right) y^{11/2}\nonumber\\
&&-\frac{856}{1575}\pi y^6\nonumber\\
&&+\left(\frac{1727123297}{604800}-\frac{267590669}{524288}\pi^2+\frac{39206}{105}\gamma+\frac{40058}{135}\ln(2)+\frac{19603}{105}\ln(y)\right.\nonumber\\
&& \left. +\frac{2916}{7}\ln(3)+\frac{26973279}{2097152}\pi^4\right) y^{13/2}\nonumber\\
&&+\frac{8636}{147}\pi y^7\nonumber\\
&&+\left(\frac{1802670191688469}{18336780000}-\frac{2224682621317007}{237817036800}\pi^2-\frac{223317}{77}\ln(3)+\frac{4484235971}{16372125}\ln(2)-\frac{3246009239}{1488375}\gamma\right. \nonumber\\
&&-\frac{3246009239}{2976750}\ln(y)-\frac{86776713349}{1610612736}\pi^4+\frac{3424}{1575}\ln(2)^2+\frac{3424}{1575}\gamma\ln(y)\nonumber\\
&& \left.+\frac{6848}{1575}\gamma\ln(2)+\frac{3424}{1575}\ln(2)\ln(y)-\frac{64}{15}\zeta(3)+\frac{856}{1575}\ln(y)^2+\frac{3424}{1575}\gamma^2\right) y^{15/2} \nonumber\\
&&+\frac{7531774}{24255}\pi y^8\nonumber\\
&&+\left(\frac{14620969720956075313}{14562737280000}+\frac{841954257}{254800}\ln(3)-\frac{2607400020943}{1833678000}\ln(y)-\frac{4849325459983}{154791000}\ln(2)\right.\nonumber\\
&& -\frac{2669989563343}{916839000}\gamma+\frac{1005859375}{185328}\ln(5)+\frac{269728}{315}\zeta(3)-\frac{15577328}{33075}\gamma^2-\frac{62406896}{33075}\ln(2)^2\nonumber\\
&&-\frac{3894332}{33075}\ln(y)^2-\frac{2970208}{1575}\gamma\ln(2)-\frac{1485104}{1575}\ln(2)\ln(y)-\frac{15577328}{33075}\gamma\ln(y)\nonumber\\
&&\left. -\frac{2756098577896711}{369937612800}\pi^2-\frac{7228107040550893}{773094113280}\pi^4\right) y^{17/2}\nonumber\\
&&+\left(-\frac{21795132889051}{9833098275}\pi+\frac{366368}{165375}\gamma\pi+\frac{366368}{165375}\pi\ln(2)-\frac{3424}{4725}\pi^3+\frac{183184}{165375}\pi\ln(y)\right) y^9\nonumber\\
&&+\left(-\frac{415840188382430911927073}{465192079672320000}+\frac{148196991257948671}{786647862000}\ln(2)+\frac{11975992669593}{134884750}\ln(3)\right.\nonumber\\
&&+\frac{3145694346466161199}{131941395333120}\pi^4 -\frac{62764648557015325903}{1171962357350400}\pi^2-\frac{1101472088331}{1073741824}\pi^6\nonumber\\
&& -\frac{4382214476}{1091475}\gamma^2-\frac{2166015625}{30888}\ln(5)-\frac{12928086}{2695}\ln(3)^2\nonumber\\
&&+\frac{55360630827270719}{1573295724000}\ln(y)+\frac{875944}{63}\zeta(3)-\frac{1597686092}{1091475}\ln(2)^2\nonumber\\
&& +\frac{55642565421011519}{786647862000}\gamma-\frac{1095553619}{1091475}\ln(y)^2-\frac{25856172}{2695}\ln(2)\ln(3)-\frac{25856172}{2695}\gamma\ln(3)\nonumber\\
&& \left.-\frac{2302692232}{363825}\gamma\ln(2)-\frac{1151346116}{363825}\ln(2)\ln(y)-\frac{12928086}{2695}\ln(3)\ln(y)-\frac{4382214476}{1091475}\gamma\ln(y)\right) y^{19/2}\nonumber\\
&&
+O_{\ln{}}(y^{10})\,.
\end{eqnarray}

\end{widetext}

\section{Eigenvalues}

We compute below the 1SF contribution to the eigenvalues of the tidal-electric, and tidal-magnetic, quadrupolar tensors $m_2^2{\mathcal E}^\mu{}_\nu(U)$,  $m_2^2{\mathcal B}^\mu{}_\nu(U)$. These eigenvalues are such that 
\begin{eqnarray}
\label{eq:4.1}
m_2^2 {\mathcal E}(U)&=& {\rm diag} [\lambda_1^{\rm (E)},\lambda_2^{\rm (E)},-(\lambda_1^{\rm (E)}+\lambda_2^{\rm (E)})]\nonumber\\
m_2^2 {\mathcal B}(U)&=& {\rm diag} [\lambda^{\rm (B)},-\lambda^{\rm (B)},0]\,,
\end{eqnarray}
where we used their tracelessness, and the existence of a zero eigenvalue of ${\mathcal B}(U)$ \cite{Dolan:2014pja}. Let us introduce a notation for the eigenvalues of the corresponding rescaled tidal tensors
\begin{eqnarray}
\label{eq:4.2}
m_2^2 {\mathcal E}(k)&=& {\rm diag} [\sigma_1^{\rm (E)},\sigma_2^{\rm (E)},-(\sigma_1^{\rm (E)}+\sigma_2^{\rm (E)})]\nonumber\\
m_2^2 {\mathcal B}(k)&=& {\rm diag} [\sigma^{\rm (B)},-\sigma^{\rm (B)},0]\,,
\end{eqnarray} 
evaluated with respect to $k$ instead of $U$.
The two set of eigenvalues are related by
\begin{eqnarray}
\label{eq:4.5}
\lambda_a^{\rm (E)}=\Gamma^2 \sigma_a^{\rm (E)}\,,\qquad
\lambda^{\rm (B)}=\Gamma^2 \sigma^{\rm (B)}\,,
\end{eqnarray}
where the 1SF expansion of the redshift factor
\beq
\Gamma= \frac{1}{\sqrt{1-3y}}-q\frac{z_1^{\rm 1SF}(y)}{1-3y}\,,
\eeq
has been derived to first order in spin in our previous work \cite{Bini:2018zde}.
One finds
\begin{eqnarray}
\lambda^{\rm (E)}_1&=& \lambda^{\rm (E)\, 0SF}_1 +q \lambda^{\rm (E)\, 1SF}_1\,,\nonumber\\
\lambda^{\rm (E)}_2&=& \lambda^{\rm (E)\, 0SF}_2 +q \lambda^{\rm (E)\, 1SF}_2\,,\nonumber\\
\lambda^{\rm (B)}&=& \lambda^{\rm (B)\, 0SF} +q \lambda^{\rm (B)\, 1SF}\,,
\end{eqnarray}
where the unperturbed (0SF) values of these eigenvalues are given by\footnote{
These eigenvalues have been already computed in Ref. \cite{Bini:2015kja} (see Eq. (4.44) there), but appear misprinted.
}
\begin{eqnarray}
\lambda^{\rm (E)\, 0SF}_1&=&  -y^3 \frac{2-3 y}{1-3y}-3\hat s y^{9/2}\frac{2-5 y}{1-3y}\,,\nonumber\\
\lambda^{\rm (E)\, 0SF}_2&=&  \frac{y^3}{1-3y}+3\hat s y^{9/2}\frac{1-2y}{1-3y}\,,\\
-\lambda^{\rm (B)\, 0SF}&=& -3 y^{7/2}\frac{\sqrt{1-2 y}}{1-3y}-\hat s\frac{3y^5 (2-5 y)}{\sqrt{1-2y}(1-3y)}\,,\nonumber
\end{eqnarray}
and the 1SF corrections are
\begin{eqnarray}
\lambda^{\rm (E)\, 1SF}_{1}&=&  \lambda^{\rm (E)\, 1SF}_{1 \, \hat s^0} + \hat s  \lambda^{\rm (E)\, 1SF}_{1 \, \hat s^1}\,,\nonumber\\
\lambda^{\rm (E)\, 1SF}_{2}&=&  \lambda^{\rm (E)\, 1SF}_{2 \, \hat s^0} + \hat s  \lambda^{\rm (E)\, 1SF}_{2 \, \hat s^1}\,,\nonumber\\
\lambda^{\rm (B)\, 1SF}&=&  \lambda^{(B)\, 1SF}_{\hat s^0} + \hat s  \lambda^{\rm (B)\, 1SF }_{\hat s^1}\,.
\end{eqnarray}

By introducing the notation \cite{Bini:2014zxa}
\begin{eqnarray}
\label{eq:4.6}
\alpha_{\rm 1SF}&=&\frac12 \tilde {\mathcal J}_{e^2}^{(0)}\delta_{e^2}(y)\,, \nonumber\\
\beta_{\rm 1SF}&=& \frac13 \tilde {\mathcal J}_{e^3}^{(0)}\delta_{e^3}(y)\,,
\end{eqnarray}
the 1SF perturbation of the exact equations
\begin{eqnarray}
\label{eq:4.8}
\frac12 m_2^4 {\rm Tr}[{\mathcal E}^2(k)]&=&\sigma_1^{\rm (E)}{}^2+\sigma_2^{\rm (E)}{}^2+\sigma_1^{\rm (E)}\sigma_2^{\rm (E)}\,,\nonumber\\
\frac13 m_2^6  {\rm Tr}[{\mathcal E}^3(k)]&=& -\sigma_1^{\rm (E)} \sigma_2^{\rm (E)}(\sigma_1^{\rm (E)}+\sigma_2^{\rm (E)})\,,
\end{eqnarray} 
yields a linear system of two equations for the two unknowns $\sigma_1^{\rm (E)1SF}$, $\sigma_2^{\rm (E)1SF}$ with $\alpha_{\rm 1SF}$ and $\beta_{\rm 1SF}$ as right hand sides.
The (unique) solution of this system reads
\begin{eqnarray}
\label{eq:4.9}
 \sigma_1^{\rm (E) 1SF}  &=& \frac{\alpha_{\rm 1SF}  \sigma_1^{\rm (E)\, 0SF} +\beta_{\rm 1SF} }{ ( \sigma_1^{\rm (E)\, 0SF}-  \sigma_2^{\rm (E)\, 0SF})
(2 \sigma_1^{\rm (E)\, 0SF}+   \sigma_2^{\rm (E)\, 0SF})}\,, \nonumber\\
\sigma_2^{\rm (E) 1SF}  &=& \frac{\alpha_{\rm 1SF}  \sigma_2^{\rm (E)\, 0SF} +\beta_{\rm 1SF} }{ ( \sigma_2^{\rm (E)\, 0SF}-  \sigma_1^{\rm (E)\, 0SF})
(2  \sigma_2^{\rm (E)\, 0SF}+   \sigma_1^{\rm (E)\, 0SF})}\,.\nonumber\\
\end{eqnarray}
As already discussed in Ref. \cite{Bini:2014zxa}, we recall that the denominators $(2 \sigma_1^{\rm (E)\, 0SF}+   \sigma_2^{\rm (E)\, 0SF})$ and $(2  \sigma_2^{\rm (E)\, 0SF}+   \sigma_1^{\rm (E)\, 0SF})$ have different PN orders. Indeed, in the Newtonian limit ($y\to 0$) $\sigma_1^{\rm (E)\, 0SF}\simeq -2y^3$ and $\sigma_2^{\rm (E)\, 0SF}\simeq +y^3$, so that $(2 \sigma_1^{\rm (E)\, 0SF}+   \sigma_2^{\rm (E)\, 0SF})\simeq -3y^3$, while $(2  \sigma_2^{\rm (E)\, 0SF}+   \sigma_1^{\rm (E)\, 0SF})=O(y^4)$ is of 1PN fractional magnitude.
As a consequence, one PN level in the analytic accuracy of $\sigma_2^{\rm (E)\, 0SF}$ is lost. 

The 1SF correction to the tidal-magnetic eigenvalue $\lambda^{\rm (B)}$ is simply evaluated as
\beq
\sigma^{\rm (B)\, 1SF}=\frac{\tilde {\mathcal J}_{b^2}^{(0)}\delta_{b^2}(y)}{4\sigma^{{\rm (B)}\, 0SF}}\,.
\eeq

The regularized value of the zeroth order in spin corrections of the $U$-normalized eigenvalues $\lambda^{\rm (E)\, 1SF}_{1 \, \hat s^0}$, $\lambda^{\rm (E)\, 1SF}_{2 \, \hat s^0}$ and $\lambda^{\rm (B)\, 1SF}_{\hat s^0}$ are known \cite{Bini:2014zxa,Dolan:2014pja,Kavanagh:2015lva,Nolan:2015vpa}.
We recall below for completeness only the first few PN terms

\begin{widetext}

\begin{eqnarray}
\lambda^{\rm (E)\, 1SF}_{1 \, \hat s^0} &=& 2 y^3+2 y^4-\frac{19}{4}y^5+\left(\frac{227}{3}-\frac{593}{256}\pi^2\right) y^6\nonumber\\
&&+\left(-\frac{71779}{4800}-\frac{719}{256}\pi^2+\frac{1536}{5}\ln(2)+\frac{384}{5}\ln(y)+\frac{768}{5}\gamma\right)y^7
+O_{{\rm ln}}(y^8)
\,,\nonumber\\
\lambda^{\rm (E)\, 1SF}_{2 \, \hat s^0}&=&  -y^3-\frac32 y^4-\frac{23}{8}y^5+\left(-\frac{2593}{48}+\frac{1249}{1024}\pi^2\right) y^6\nonumber\\
&& +\left(-\frac{362051}{3200}-\frac{128}{5}\ln(y)+\frac{1737}{1024}\pi^2-\frac{256}{5}\gamma-\frac{512}{5}\ln(2)\right) y^7
+O_{{\rm ln}}(y^8)
\,,\nonumber\\
-\lambda^{\rm (B)\, 1SF}_{\hat s^0}&=&  
2y^{7/2}+3y^{9/2}+\frac{59}{4}y^{11/2}+\left(\frac{2761}{24}-\frac{41}{16}\pi^2\right)y^{13/2}\nonumber\\
&&+\left(\frac{1618039}{2880}-\frac{112919}{3072}\pi^2+\frac{1808}{15}\gamma+240\ln(2)+\frac{904}{15}\ln(y) \right)y^{15/2}
+O_{{\rm ln}}(y^{17/2})
\,.
\end{eqnarray}

The corrections linear in spin are given by 

\begin{eqnarray}
\lambda^{\rm (E)\, 1SF}_{1 \, \hat s^1}&=&  
8 y^{9/2}-12 y^{11/2}-18 y^{13/2}+\left(\frac{659}{3}-\frac{2435}{256}\pi^2\right) y^{15/2}\nonumber\\
&+& \left(\frac{902941}{3600}-\frac{154267}{3072}\pi^2+\frac{752}{3}\gamma+\frac{2448}{5}\ln(2)+\frac{376}{3}\ln(y)\right) y^{17/2}\nonumber\\
&+& \left(\frac{55233761}{16800}-\frac{1272409}{8192}\pi^2-\frac{59176}{35}\gamma-\frac{29588}{35}\ln(y)-\frac{319352}{105}\ln(2)-\frac{2187}{7}\ln(3)\right) y^{19/2}\nonumber\\
&+&\frac{43656}{175}\pi y^{10}\nonumber\\
&+&\left(-\frac{63262994029}{793800}+\frac{71412166381}{7077888}\pi^2-\frac{443678}{405}\gamma-\frac{19495058}{2835}\ln(2)-\frac{221839}{405}\ln(y)+\frac{89667}{70}\ln(3)\right.\nonumber\\
&& \left. -\frac{60882449}{1048576}\pi^4\right) y^{21/2}\nonumber\\
&-&\frac{18887593}{11025}\pi y^{11}\nonumber\\
&&
+\left(\frac{77805223927}{1819125}\gamma-\frac{3105568}{1575}\gamma^2+\frac{672258572747}{402653184}\pi^4+\frac{710346154789}{5457375}\ln(2)+\frac{16466409}{6160}\ln(3)\right.\nonumber\\
&&
-\frac{2461856}{315}\ln(2)^2-\frac{224609375}{14256}\ln(5)+\frac{58048}{15}\zeta(3)+\frac{897434891715947}{19818086400}\pi^2-\frac{4115648}{525}\gamma\ln(2)\nonumber\\
&&\left.
-\frac{9073265747546491}{13752585000}+\frac{77805223927}{3638250}\ln(y)-\frac{3105568}{1575}\gamma\ln(y)-\frac{2057824}{525}\ln(2)\ln(y)-\frac{776392}{1575}\ln(y)^2\right)y^{23/2}\nonumber\\
&&
-\frac{1560074701}{2182950}\pi y^{12}\nonumber\\
&&
+\left(-\frac{35676003719939}{361179000}\gamma+\frac{163272544}{11025}\gamma^2+\frac{676470994112801}{42949672960}\pi^4-\frac{1669871109946433}{3972969000}\ln(2)\right.\nonumber\\
&&
-\frac{21959878162869}{196196000}\ln(3)+\frac{536078176}{11025}\ln(2)^2+\frac{170586}{49}\ln(3)^2+\frac{504291015625}{2594592}\ln(5)-\frac{230784}{7}\zeta(3)\nonumber\\
&&
+\frac{1402717571812867}{10276044800}\pi^2+\frac{575082176}{11025}\gamma\ln(2)+\frac{341172}{49}\gamma\ln(3)+\frac{341172}{49}\ln(2)\ln(3)\nonumber\\
&&
-\frac{35306156423939}{722358000}\ln(y)-\frac{4153593072625400734241}{1409672968704000}+\frac{163272544}{11025}\gamma\ln(y)+\frac{170586}{49}\ln(3)\ln(y)\nonumber\\
&&\left.
+\frac{287541088}{11025}\ln(2)\ln(y)+\frac{40818136}{11025}\ln(y)^2\right)y^{25/2}
+O_{\ln{}}(y^{13})\,,
\end{eqnarray}

\begin{eqnarray}
\lambda^{\rm (E)\, 1SF}_{2 \, \hat s^1}&=&   
-4y^{9/2}+2y^{11/2}+\frac{9}{2}y^{13/2}
+\left(-\frac{1627}{12}+\frac{2561}{512}\pi^2\right)y^{15/2}\nonumber\\
&&
+\left(\frac{610013}{7200}-\frac{243815}{24576}\pi^2-\frac{2144}{15}\gamma-\frac{1072}{15}\ln(y)-\frac{4192}{15}\ln(2)\right)y^{17/2}\nonumber\\
&&
+\left(\frac{85419731}{6720}-\frac{2819407}{2048}\pi^2+\frac{20796}{35}\gamma+\frac{10398}{35}\ln(y)+\frac{19244}{15}\ln(2)-\frac{729}{7}\ln(3)\right)y^{19/2}\nonumber\\
&&
-\frac{224272}{1575}\pi y^{10}\nonumber\\
&&
+\left(\frac{2892840540493}{25401600}-\frac{374039842301}{28311552}\pi^2+\frac{515194}{405}\gamma+\frac{4197526}{2835}\ln(2)+\frac{257597}{405}\ln(y)\right.\nonumber\\
&&\left.
+\frac{43011}{35}\ln(3)+\frac{735266513}{8388608}\pi^4\right)y^{21/2}\nonumber\\
&&
+\frac{2301737}{3675}\pi y^{11}\nonumber\\
&&
+\left(-\frac{24978143909}{1559250}\gamma-\frac{24978143909}{3118500}\ln(y)+\frac{253376}{225}\gamma^2+\frac{879445089053}{1006632960}\pi^4\right.\nonumber\\
&&
-\frac{5956778549}{198450}\ln(2)-\frac{5490099}{1540}\ln(3)+\frac{7032896}{1575}\ln(2)^2+\frac{9765625}{7128}\ln(5)-\frac{33152}{15}\zeta(3)\nonumber\\
&&
-\frac{98262348401861}{1887436800}\pi^2+\frac{1410688}{315}\gamma\ln(2)+\frac{63344}{225}\ln(y)^2+\frac{54665096945177993}{125737920000}\nonumber\\
&&\left.
+\frac{705344}{315}\ln(2)\ln(y)+\frac{253376}{225}\gamma\ln(y)\right)y^{23/2}
+O_{\ln{}}(y^{12})\,,
\end{eqnarray}

\begin{eqnarray}
-\lambda^{\rm (B)\, 1SF}_{\hat s^1}&=&   
6y^5-\frac{21}{2}y^6+\left(31-\frac{189}{2048}\pi^2\right)y^7
+\left(\frac{97537}{400}-\frac{7599}{1024}\pi^2+\frac{4}{5}\ln(y)+\frac{8}{5}\ln(2)+\frac{8}{5}\gamma\right)y^8\nonumber\\
&&
+\left(\frac{321104729}{100800}-\frac{117514627}{393216}\pi^2+\frac{1234}{15}\ln(y)+\frac{2468}{15}\gamma+\frac{33916}{105}\ln(2)\right)y^9\nonumber\\
&&
+\frac{428}{525}\pi y^{19/2}\nonumber\\
&&
+\left(\frac{1154329703}{403200}+\frac{313611845}{3145728}\pi^2-\frac{91853}{105}\gamma-\frac{511207}{315}\ln(2)-\frac{91853}{210}\ln(y)-\frac{729}{7}\ln(3)-\frac{80919837}{4194304}\pi^4\right)y^{10}\nonumber\\
&&
+\frac{580942}{3675}\pi y^{21/2}\nonumber\\
&&
+\left(-\frac{342516124855213}{1811040000}+\frac{1113494091353669}{52848230400}\pi^2-\frac{550915387}{330750}\gamma-\frac{550915387}{661500}\ln(y)-\frac{15669434057}{3638250}\ln(2)\right.\nonumber\\
&&
-\frac{264141}{770}\ln(3)+\frac{32}{5}\zeta(3)-\frac{63982008699}{1073741824}\pi^4-\frac{3424}{525}\gamma\ln(2)-\frac{1712}{525}\gamma^2-\frac{1712}{525}\ln(2)^2\nonumber\\
&&\left.
-\frac{428}{525}\ln(y)^2-\frac{1712}{525}\gamma\ln(y)-\frac{1712}{525}\ln(2)\ln(y)\right)y^{11}\nonumber\\
&&
-\frac{105172666}{121275}\pi y^{23/2}\nonumber\\
&&
+\left(\frac{14401009433479}{611226000}\gamma-\frac{1534056}{1225}\gamma^2+\frac{8191460400868013}{515396075520}\pi^4+\frac{9128850893003}{162162000}\ln(2)\right.\nonumber\\
&&
+\frac{25951426299}{5605600}\ln(3)-\frac{6090856}{1225}\ln(2)^2-\frac{2001953125}{370656}\ln(5)+\frac{268784}{105}\zeta(3)\nonumber\\
&&
+\frac{189103487160367811}{2219625676800}\pi^2-\frac{2616848}{525}\gamma\ln(2)-\frac{1534056}{1225}\gamma\ln(y)-\frac{768140015394994675117}{320380220160000}\nonumber\\
&&\left.
-\frac{1308424}{525}\ln(2)\ln(y)+\frac{14338419891079}{1222452000}\ln(y)-\frac{383514}{1225}\ln(y)^2\right)y^{12}\nonumber\\
&&
+\left(-\frac{362196722747}{262215954}\pi+\frac{1712}{1575}\pi^3-\frac{183184}{55125}\gamma\pi-\frac{91592}{55125}\pi\ln(y)-\frac{183184}{55125}\pi\ln(2)\right)y^{25/2}\nonumber\\
&&
+\left(-\frac{2738651457992501}{74918844000}\gamma+\frac{3304416264993}{2147483648}\pi^6+\frac{382334818}{51975}\gamma^2+\frac{5177433837486038387}{263882790666240}\pi^4\right.\nonumber\\
&&
-\frac{8316327878531317}{74918844000}\ln(2)-\frac{10245646667673}{269769500}\ln(3)+\frac{260978042}{10395}\ln(2)^2+\frac{3755079}{2695}\ln(3)^2\nonumber\\
&&
+\frac{23521484375}{432432}\ln(5)-\frac{79044}{5}\zeta(3)+\frac{5703631259017453919}{15945066086400}\pi^2+\frac{459902188}{17325}\gamma\ln(2)\nonumber\\
&&
+\frac{7510158}{2695}\gamma\ln(3)+\frac{7510158}{2695}\ln(2)\ln(3)+\frac{382334818}{51975}\gamma\ln(y)+\frac{3755079}{2695}\ln(3)\ln(y)\nonumber\\
&&
+\frac{229951094}{17325}\ln(2)\ln(y)-\frac{2686867553019701}{149837688000}\ln(y)-\frac{14630621620415023880359}{2109714647040000}\nonumber\\
&&\left.
+\frac{191167409}{103950}\ln(y)^2\right)y^{13}
+O(y^{27/2})\,.
\end{eqnarray}

\end{widetext}

\begin{figure}
\includegraphics[scale=0.35]{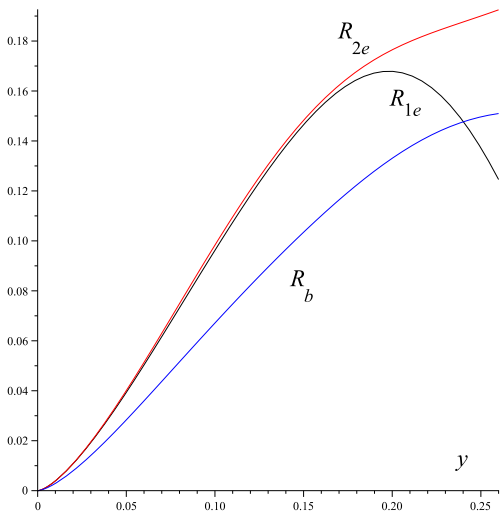}
\caption{\label{fig:1} 
The ratios $R_{1\,e}$, $R_{2\,e}$ and $R_{b}$ between the 1SF first-order in spin corrections to the electric and magnetic eigenvalues and the zeroth order ones are plotted as functions of $y$.
}
\end{figure}

We show in Fig. \ref{fig:1} the behavior of the ratios between the 1SF first-order in spin contributions and the zeroth order ones $R_{1\,e}=\lambda^{\rm (E)\, 1SF}_{1 \, \hat s^1}/\lambda^{\rm (E)\, 1SF}_{1 \, \hat s^0}$, $R_{2\,e}=\lambda^{\rm (E)\, 1SF}_{2 \, \hat s^1}/\lambda^{\rm (E)\, 1SF}_{2 \, \hat s^0}$ and $R_{b}=\lambda^{\rm (B)\, 1SF}_{\hat s^1}/\lambda^{\rm (B)\, 1SF}_{\hat s^0}$ as functions of $y$.
In the weak field region, well described by the PN approximation, their behavior is characterized by a general increasing. 
The maximum amount of these ratios is about 10\% in the range $y\in[0,0.1]$, a value which is almost doubled increasing the $y$-interval in $[0,0.2]$. Therefore, the contribution due to spin is in general smaller that $0.1\hat s$ times the zeroth order term.
Approaching the strong field region ($y\gtrapprox0.15$) we see the onset of typical PN oscillations, whose physical meaning can be inferred only by significantly raising the accuracy theoretically (to very high-PN orders as in Ref. \cite{Kavanagh:2015lva} for the spinless case) or by performing fully numerical analyses.
Unfortunately, the lack in the literature of numerical studies of these quantities prevents us to have more insight in the strong field region.


\section{Concluding remarks}

In this work we have analytically computed the 1SF corrections to the (quadratic and cubic) electric-type and (quadratic) magnetic-type tidal invariants for a binary system consisting of a nonrotating body (with larger mass $m_2$) and an extended body endowed with spin (with smaller mass $m_1$) to the first-order in spin and in the extreme-mass-ratio limit $m_1\ll m_2$, generalizing previous results, where the perturbing body was spinless.
Here, the orbit of the smaller body is still circular (or \lq\lq helical" in the perturbed spacetime having a Killing helical symmetry) but nongeodesic, due to the coupling between spin and curvature as described by the Mathisson-Papapetrou-Dixon model for spinning bodies.
Nonvanishing spin and nongeodesic orbits are then two original contributions of the present analysis to the study of two-body tidal interactions.
The inclusion of spin corrections has become possible after the work of Ref. \cite{Bini:2018zde}, where the redshift function $z_1$ of a spinning particle in a perturbed Schwarzschild spacetime was computed, once the completion problem of metric reconstruction was fully solved by the addition of \lq\lq low multipoles" $l=0,1$.

Our results are accurate through the 9.5PN order for both electric and magnetic tidal invariants. We have also computed the associated eigenvalues, whose accuracy is still 9.5PN for $\lambda^{\rm (E)\, 1SF}_{1}$ and $\lambda^{\rm (B)\, 1SF}$, whereas it is one PN order less for $\lambda^{\rm (E)\, 1SF}_{2}$.
The contribution due to spin to the 1SF corrections to tidal eigenvalues is in general about $0.1\hat s$ times the zeroth-order-in-spin term in a spacetime region where the weak-field approximation holds.
Our analytical results can be used to easily compute related gauge-invariant quantities, like the conservative part of the speciality index, whose corrections mark the change of Petrov spectral type of the (algebraically special type-D) background metric in an invariant way \cite{Cherubini:2004yi,Cherubini:2004au,Dolan:2014pja}. 
This information can then be useful to test numerical relativity results in the extreme-mass-ratio limit. 

Another important task will be the conversion of these results into the EOB formalism. 
Actually, following Ref. \cite{Bini:2012gu} spin modification to the tidal potentials should naturally enter the EOB main radial potential $A(u)$ (see Eq. (1.1) there)
\beq
A(u)=A^{\rm BBH}(u)+A^{\rm tidal}_1(u)+A^{\rm tidal}_2(u)\,,
\eeq 
where $A^{\rm BBH}(u)$ is the potential describing the dynamics of the binary system and $A^{\rm tidal}_{1,2}(u)$ those associated with the tidal deformations of the two bodies.
Spin corrections can thus be included in $A^{\rm tidal}_{1,2}(u)$, with for instance
\beq
A^{\rm tidal}_{1}(u)=A^{{\rm tidal}\, S_{1}^0}_{1}(u)+S_{1}A^{{\rm tidal}\, S_{1}^1}_{1}(u)+O(S_1^2)\,,
\eeq
where $S_{1}$ denotes the spin of the body 1.
However, linear-in-spin terms naturally enter also the spin-orbit part of the EOB Hamiltonian, providing a tidal modification of the two gyrogravitomagnetic ratios $g_S$ and $g_{S^*}$. Indeed, the radial potential $A(u)$ incorporates typically even-in-spin corrections, whereas $g_S$ and $g_{S^*}$ include odd-in-spin corrections.
This problem is currently under investigation and will be addressed elsewhere.

\section*{Acknowledgments}
The authors thank T. Damour for useful discussions.
D.B. thanks the Naples Section of the Italian Istituto Nazionale di Fisica Nucleare (INFN) and the International Center for Relativistic Astrophysics Network (ICRANet) for partial support.

\end{document}